\documentclass[12pt]{article}
\usepackage{amsmath,amssymb,cite}

\def\be{\begin{equation}}
\def\eqn{\begin{equation}\label}
\newcommand{\ee}{\end{equation}}

\def\bea{\begin{eqnarray}}
\def\eqnn{\bea\label}
\def\eea{\end{eqnarray}}
\def\nn{\nonumber}

\def\md{\medskip}

\newcommand{\eqna}[1]{\begin{subequations} \label{#1}
\begin{eqnarray}}
\def\eena{\end{eqnarray}
\end{subequations}}

 \def\nt{\noindent}

\font\fat=cmsy10 scaled\magstep5

\def\Bbullet{\raise-3pt\hbox{\fat\char"0F}}

\def\tV{\tilde V}

\def\bu{$\bullet$}
\def\buu{\bullet\ }

\def\cg{{\cal G}} \def\ch{{\cal H}}

\def\riga{-\kern-4pt - \kern-4pt -}

\def\bbc{C\kern-6pt I}
\def\bac{{C\kern-5.5pt I}}

 \def\k{\kappa} \def\o{{\bar 0}} \def\I{{\bar 1}}

\def\a{\alpha}   \def\d{\delta}

\def\D{\Delta} \def\L{\Lambda}

\def\Ra{\Longrightarrow}

\def\({\left(}
\def\){\right)}

\def\a{\alpha}

\def\d{\delta}

\def\D{\Delta}
\def\L{\Lambda}

\begin{document}

\begin{center}

{\LARGE {\bf Group-Theoretical
 Classification\\ of BPS  States in\\ D=4
  Conformal Supersymmetry:\\[2pt]
  the Case of \boldmath{$\frac{1}{N}\,$}-BPS}
\footnote{Plenary talk at the International Workshop
'Supersymmetries and Quantum Symmetries', Dubna, 18-23.7.2011.}}

 \vspace{10mm}

{\bf \large V.K. Dobrev}

\vskip 5mm

\emph{Institute for Nuclear
Research and Nuclear Energy, \\
 Bulgarian Academy of Sciences,\\ 72
Tsarigradsko Chaussee, 1784 Sofia, Bulgaria}

\end{center}

\vskip 4mm

\begin{abstract}
{\small In an earlier  paper  we gave the complete group-theoretical
classification of BPS states of the N-extended D=4 conformal
superalgebras su(2,2/N), but not all interesting cases were given in
detail. In the present paper  we spell out the interesting case of
~$\frac{1}{N}$~-BPS~    and
   possibly protected states. }
\end{abstract}

\vskip 10mm

\section{Introduction}
\setcounter{equation}{0}

Recently, superconformal field theories in various dimensions are
attracting more interest, especially in view of their applications in string theory.
Thus,  the classification of the UIRs of the conformal
 superalgebras  is  of great importance.
For some time  such classification was known only for the ~$D=4$~
superconformal algebras ~$su(2,2/1)$ \cite{FF} and ~$su(2,2/N)$~ for
arbitrary $N$ \cite{DPu}, (see also \cite{DPm,DPf}).
 Then, more progress was made with the classification
for ~$D=3$ (for even $N$), $D=5$, and $D=6$ (for $N=1,2$)~
in \cite{Min} (some results being conjectural), then for the $D=6$
case (for arbitrary $N$) was finalized in \cite{Dosix}. Finally, the
cases $D=9,10,11$ were treated by finding the UIRs of $osp(1/2n)$,
\cite{DoZh}.

After we have know the UIRs the next problem to address is to find
their characters since these give the spectrum which is important
for the applications. This was done for the UIRs of $D=4$ conformal
superalgebras $su(2,2/N)$ in \cite{Doch}. From the mathematical
point of view this question is clear only for representations with
conformal dimension above the unitarity threshold viewed as irreps
of the corresponding complex superalgebra $sl(4/N)$.  But for
$su(2,2/N)$ even the UIRs above the unitarity threshold are
truncated for small values of spin and isospin. More than that, in
the applications the most important role is played by the
representations with ``quantized" conformal dimensions at the
unitarity threshold and at discrete points below.  In the quantum
field or string theory framework some of these correspond to
operators with ``protected" scaling dimension and therefore imply
``non-renormalization theorems" at the quantum level, cf., e.g.,
\cite{HeHo,FSa}. Especially important in this context are the
so-called BPS states, cf.,
\cite{AFSZ,FSb,FSa,EdSo,Ryz,DHRy,ArSo,DHHR}.

These investigations requires deeper knowledge of the structure of the UIRs.    Fortunately,
most of the needed  information is contained in \cite{DPu,DPm,DPf,DPp}.
We use also more explicit results on the decompositions of long superfields
as they descend to the unitarity threshold \cite{Doch}.

In the paper \cite{dob-bps} we gave the complete group-theoretical
classification of the BPS states, but not all interesting cases were
given in detail. In the present paper motivated by the paper
\cite{BHSV} we spell out the interesting case of
~$\frac{1}{N}$~-BPS~ states.

\section{Preliminaries}
\setcounter{equation}{0}

\subsection{Representations of D=4 conformal supersymmetry}

 The conformal superalgebras in $D=4$ are ~$\cg ~=~ su(2,2/N)$.
The even subalgebra of ~$\cg$~ is the algebra ~$\cg_0 ~=~ su(2,2)
\oplus u(1) \oplus su(N)$. We label their physically relevant
representations of ~$\cg$~ by the signature: \eqn{sgn}\chi ~=~
[\,d\,;\,
j_1\,,\,j_2\,;\,z\,;\,r_1\,,\ldots,r_{N-1}\,]\end{equation}  where
~$d$~ is the conformal weight, ~$j_1,j_2$~ are non-negative
(half-)integers which are Dynkin labels of the finite-dimensional
irreps of the $D=4$ Lorentz subalgebra ~$so(3,1)$~ of dimension
~$(2j_1+1)(2j_2+1)$, ~$z$~ represents the ~$u(1)$~ subalgebra which
is central for ~$\cg_0$~ (and  is central for $\cg$ itself when
$N=4$), and ~$r_1,\ldots,r_{N-1}$~ are non-negative integers which
are Dynkin labels of the finite-dimensional irreps of the internal
(or $R$) symmetry algebra ~$su(N)$.

We  recall the root system of the complexification ~$\cg^\bac$~ of
$\cg$  (as used in \cite{DPf}). The positive root system ~$\D^+$~ is
comprised of ~$\a_{ij}\,$, ~$1\leq i <j \leq 4+N$. The even positive
root system ~$\D^+_\o$~ is comprised of ~$\a_{ij}\,$, with\ $i,j\leq
4$~ and ~$i,j\geq 5$; ~the odd positive root system ~$\D^+_\I$~ is
comprised of ~$\a_{ij}\,$, with ~$i\leq 4, j \geq 5$. The generators
corresponding to the latter (odd) roots will be denoted as
~$X^+_{i,4+k}\,$, where ~$i=1,2,3,4$, ~$k=1,\ldots,N$.

We use lowest weight Verma modules $V^\L$ over $\cg^\bac$, where the
lowest weight $\L$ is characterized by its values on the Cartan
subalgebra ~$\ch$~ and is in 1-to-1 correspondence with the
signature $\chi$. If a Verma module ~$V^\L$~ is irreducible then it
gives the lowest weight irrep ~$L_\L$~ with the same weight. If a
Verma module ~$V^\L$~ is reducible then it contains a maximal
invariant submodule ~$I^\L$~ and the lowest weight irrep ~$L_\L$~
with the same weight is given by factorization: ~$L_\L ~=~
V^\L\,/\,I^\L$ \cite{Kacl}.

There are submodules which are generated by the singular vectors
related to the even simple roots ~\cite{DPf}. These generate an even
invariant submodule $I^\L_c$ present in all Verma modules that we
consider and which must be factored out. Thus, instead of ~$V^\L$~
we shall consider the factor-modules: \eqn{fcc} \tV^\L ~=~ V^\L\,
/\, I^\L_c
\end{equation}

The Verma module reducibility conditions for the ~$4N$~ odd positive
roots of $\cg^\bac$ were derived in \cite{DPm,DPf} adapting the
results of Kac \cite{Kacl}: \eqna{redu} && d ~=~ d^1_{Nk} - z
\d_{N4}
\\ &&  d^1_{Nk} ~\equiv~ 4-2k +2j_2 +z+2m_k -2m/N \cr
&&\cr && d ~=~ d^2_{Nk} - z \d_{N4} \\ &&  d^2_{Nk} ~\equiv~ 2-2k
-2j_2 +z+2m_k -2m/N
\cr &&\cr && d ~=~ d^3_{Nk} + z \d_{N4}\\
&& d^3_{Nk} ~\equiv~ 2+2k-2N +2j_1 -z-2m_k +2m/N \cr &&\cr && d ~=~
d^4_{Nk} + z \d_{N4}\\ && d^4_{Nk} ~\equiv~ 2k-2N -2j_1 -z-2m_k
+2m/N \nn\eena \nt where in all four cases of (\ref{redu})
~$k=1,\ldots,N$, ~$m_N\equiv 0$, and \eqn{mkm} m_k \equiv
\sum_{i=k}^{N-1} r_i \ , \quad m \equiv \sum_{k=1}^{N-1} m_k =
\sum_{k=1}^{N-1} k r_k\end{equation} Note that we shall use also the
quantity $m^*$ which is conjugate to $m$~:
\eqnn{conjm}  m^* ~&\equiv&~ \sum_{k=1}^{N-1} k r_{N-k} ~=~ \sum_{k=1}^{N-1} (N-k) r_{k}\ ,\\
&& m+m^* ~=~ Nm_1 \ .\eea

We need  the result of \cite{DPu} (cf. part (i) of the Theorem
there) that the following is the complete list of lowest weight
(positive energy) UIRs of $su(2,2/N)$~: \eqna{unitt} && d ~\geq~
d_{\rm max} ~=~ \max (d^1_{N1}, d^3_{NN})\ ,\\ && d ~=~ d^4_{NN}
\geq d^1_{N1}\ , ~~ j_1=0\ ,\\ && d ~=~ d^2_{N1} \geq d^3_{NN}\ , ~~
j_2=0\ ,\\ && d ~=~ d^2_{N1} = d^4_{NN}\ , ~~ j_1 = j_2=0\ ,\eena
\nt where ~$d_{\rm max}$~ is the threshold of the continuous unitary
spectrum. Note that in case (d) we have $d=m_1$, $z=2m/N -m_1\,$,
and that it is trivial for $N=1$.

Next we note that if ~$d ~>~ d_{\rm max}$~ the factorized Verma
modules are irreducible and coincide with the UIRs ~$L_\L\,$. These
UIRs are called ~${\bf long}$~ in the modern literature, cf., e.g.,
\cite{FGPW,FS,FSa,AES,BKRSa,EdSo,HeHoa}. Analogously, we shall use
for the cases when ~$d= d_{\rm max}\,$, i.e., (\ref{unitt}a), the
terminology of ~{\bf semi-short}~ UIRs, introduced in
\cite{FGPW,FSa}, while the cases (\ref{unitt}{b,c,d}) are also
called ~{\bf short}~ UIRs, cf., e.g.,
\cite{FS,FSa,AES,BKRSa,EdSo,HeHoa}.

Next consider in more detail the UIRs at the four distinguished
reducibility points determining the UIRs list above: ~$d^1_{N1}\,$,
~$d^2_{N1}\,$, ~$d^3_{NN}\,$, ~$d^4_{NN}\,$.   We note a partial
ordering of these four points: \eqn{parto} d^1_{N1} ~>~ d^2_{N1} \ ,
\qquad d^3_{NN} ~>~ d^4_{NN} \ .\end{equation}
 Due to this ordering ~{\it at most two}~ of these four points may coincide.

First we consider the situations in which ~{\it no two}~ of the
distinguished four points coincide. There are four such situations:
\eqna{dist} &{\bf a:} & ~~~
 d = d_{\rm max} = d^1_{N1} =
d^a \equiv 2 +2j_2 +z+2m_1 -2m/N > d^3_{NN}\qquad\\
& {\bf b:} &  ~~~d
~=~ d^2_{N1} = d^b \equiv z -2j_2 +2m_1 -2m/N  > d^3_{NN}\ , ~~ j_2=0\  \\
& {\bf c:} &~~~
 d ~=~ d_{\rm max} ~=~
d^3_{NN} ~=~ d^c ~\equiv~ 2+2j_1 -z +2m/N  > d^1_{N1}\  \\
&{\bf d:} &~~~
 d ~=~ d^4_{NN} = d^d \equiv 2m/N-2j_1 -z  > d^1_{N1}\ , ~~ j_1=0\
\eena \noindent where for future use we have introduced notations
~$d^a,d^b,d^c,d^d$, the definitions including also the corresponding
inequality.

We shall call these cases ~{\bf single-reducibility-condition
(SRC)}~ Verma modules or UIRs, depending on the context. In
addition, as already stated, we use for the cases when ~$d= d_{\rm
max}\,$, i.e., (\ref{dist}a,c), the terminology of semi-short UIRs,
while the cases (\ref{dist}b,d), are also called short UIRs.

The factorized Verma modules ~$\tV^\L$~ with the unitary signatures
from (\ref{dist}) have only one invariant odd submodule which has to
be factorized in order to obtain the UIRs.

We consider  now the four situations in which ~{\it two}~
distinguished points coincide: \eqna{disd} &{\bf ac:} &\quad d ~=~
d_{\rm max} ~=~ d^{ac} ~\equiv~ 2 + j_1 + j_2 + m_1 =
d^1_{N1} = d^3_{NN}\\
&{\bf ad:} &\quad  d ~=~ d^{ad} ~\equiv~  1 + j_2 + m_1 ~=~ d^1_{N1}
=
d^4_{NN}  \ , ~~ j_1=0 \qquad\\
&{\bf bc:}
&\quad  d ~=~ d^{bc} ~\equiv ~ 1 + j_1 + m_1 ~=~ d^2_{N1} = d^3_{NN}  \ , ~~ j_2=0 \\
&{\bf bd:} &\quad  d ~=~ d^{bd} ~\equiv ~ m_1 ~=~ d^2_{N1} =
d^4_{NN}  \ , ~~ j_1=j_2=0 \eena \nt We shall call these ~{\bf
double-reducibility-condition (DRC)} Verma modules or UIRs. The
cases in (\ref{disd}a) are semi-short UIR, while the other cases are
short.

\section{BPS and possibly protected states}
\setcounter{equation}{0}

BPS states   are characterized by the number ~$\k$~ of odd
generators which annihilate them - then the corresponding state is
called ~${\k\over 4N}\,$-BPS state. The most interesting case for
BPS states is when ~$N=4$ since is related to super-Yang-Mills, cf.,
\cite{AFSZ,FSb,FSa,EdSo,Ryz,DHRy,ArSo,DHHR,Dolan}. Also
group-theoretically the case $N=4$  is special since the ~$u(1)$~
subalgebra carrying the quantum number $z$ becomes central and one
can invariantly set ~$z=0$. When $N\neq 4$ we can also set $z=0$
though this does not have the same group-theoretical meaning as for
$N=4$.

In the paper \cite{dob-bps} we gave the complete classification of
the BPS states, but not all interesting cases were given in detail.
In the present paper motivated by the paper \cite{BHSV} we spell out
the interesting case of ~$\frac{1}{N}$~-BPS~ states, i.e., the cases
when ~$\k=4$.

It is convenient to consider the case of general $N$ while treating
separately $R$-symmetry scalars and $R$-symmetry non-scalars.

\subsection{R-symmetry scalars}
\label{rsymscal}

We start with the simpler cases of $R$-symmetry scalars when
~$r_i=0$~  for all $i$, which means also that ~$m_1=m=m^*=0$.

These cases are valid also for $N=1$, however for ~$N=1$~ in all
cases we have ~$\k<4$, \cite{dob-bps}.

In fact only three cases are relevant for ~$\k=4$.

\nt\bu{\bf a} ~~~ $d ~=~ (d^1_{N1})_{\vert_{m=0=z}}  ~=~ 2 +2j_2 ~>~
2 +2j_1  ~=~ (d^3_{NN})_{\vert_{m=0=z}}\,$. ~~The last inequality
leads to the restriction: ~$j_2>j_1\,$, i.e., ~$j_2>0$, and then we
have: \eqn{bpsa} \k=N, ~~~~m_1=m=0, ~~~j_2>0 \
.\ee These semi-short UIRs may be called semi-chiral since they lack
half of the anti-chiral generators: ~$X^+_{3,4+k}$, $k=1,\ldots,N$.

\nt\bu{\bf c} ~~~ $d ~=~   (d^3_{NN})_{\vert_{m=0=z}} ~=~ 2 +2j_1
~>~ ~=~ (d^1_{N1})_{\vert_{m=0=z}}  ~~~\Ra$  \eqn{bpsc} \k=N, ~~
~m_1=m=0, ~j_1>0\ . \ee These semi-short UIRs may be called
semi--anti-chiral since they lack half of the chiral generators:
~$X^+_{1,4+k}$, $k=1,\ldots,N$.

Thus, in both cases above the interesting case ~$\k=4$~ occurs only
for ~$N=4$, as ~$\frac{1}{4}\,$-BPS.

\eqnn{rsymac} \buu{\bf ac} &&  d ~=~ d^{ac}_{\vert_{m=0}}~=~  2 +
j_1 + j_2 \ ,~~z ~=~ j_1-j_2\ ,\nn\\
&&\k=2N,  ~{\rm if} ~~ j_1,j_2>0 ,\nn\\  
&&\k=N+1,  ~{\rm if} ~~ j_1 >0,\ j_2=0 ,\nn\\
&&\k=N+1,  ~{\rm if} ~~ j_1=0,\ j_2>0 ,\nn\\
&&\k=2,  ~{\rm if} ~~ j_1=j_2=0 .\nn\eea Here, ~$\k$~ is the number
of mixed elimination: chiral generators ~$X^+_{1,4+k}$,
($k=1,\ldots,N + (1-N)\d_{j_1,0}$), and anti-chiral generators
~$X^+_{3,5+N-k}$, ($k=1,\ldots,N + (1-N)\d_{j_2,0}$). Thus, in the
cases when ~$\k=2N$~ the semi-short UIRs may be called
semi--chiral--anti-chiral since they lack half of the chiral and
half of the anti-chiral generators.\\
The interesting case ~$\k=4$~ occurs only for ~$N=2$, as ~$\frac{1}{2}\,$-BPS.

\subsection{R-symmetry non-scalars}
\label{rsymscal}

 Below we need some additional notation. Let $N>1$
and let ~$i_0$~ be an integer such that ~$0\leq i_0\leq N-1\,$,
~$r_i=0$~ for ~$i\leq i_0\,$, and if ~$i_0<N-1$~ then ~$r_{i_0+1}>
0$.  Let now ~$i'_0$~ be an integer such that ~$0\leq i'_0\leq
N-1\,$, ~$r_{N-i}=0$~ for ~$i\leq i'_0\,$, and if ~$i'_0<N-1$~ then
~$r_{N-1-i'_0}> 0$.

The interesting cases of ~$\frac{1}{N}$~-BPS~ states, i.e., when
~$\k=4$, are given in the following list:

\eqnn{zrsymna} \buu{\bf a} && d ~=~  d^a = 2 +2j_2 + 2m^*/N ,
~~~N\geq 5, \\ && j_1 ~{\rm arbitrary}, ~~j_2>0, ~~i_0=3, ~~0\leq
i'_0\leq N-5, \nn\\ && j_2 > j_1 + \sum_{k=4}^{N-1} (2k/N -1)r_k\
.\nn \eea Here are eliminated four anti-chiral generators
~$X^+_{3,4+k}\,$, $k\leq 4 \,$.

\eqnn{zrsymnb} \buu{\bf b} && d ~=~ d^b =  2m^*/N, ~~~N\geq 5,
\\ && ~~ j_2=0\ , ~~j_1 ~{\rm arbitrary}, ~~i_0=1, ~~0\leq
i'_0\leq N-3, \nn\\ &&  \nn\\ && \sum_{k=2}^{[(N-1)/2]} (1-2k/N)r_k
~> j_1 + \sum_{[(N+1)/2]}^{N-1} (2k/N -1)r_k\   \ .\nn\eea Here are
eliminated four anti-chiral generators ~$X^+_{3,5+N-k}\,$,
~$X^+_{4,5+N-k}\,$, $k\leq 2  \,$.

\eqnn{zrsymnc} \buu{\bf c} && d ~=~  d^c = 2 +2j_1 + 2m/N , ~~~N\geq
5, \\ && j_1 >0, ~~j_2 ~{\rm arbitrary},  ~~i'_0=3, ~~0\leq i_0\leq
N-5, \nn\\ && j_1 > j_2 + \sum_{k=1}^{N-4} (1-2k/N)r_k\ .\nn \eea
 Here are eliminated four chiral generators  ~$X^+_{1,4+k}\,$, $k\leq 4 \,$.

\eqnn{zrsymnd} \buu{\bf d}  && d ~=~ d^d =  2m/N, ~~~N\geq 5,
\\ && ~~ j_1=0\ , ~~j_2 ~{\rm arbitrary}, ~~i'_0=1, ~~0\leq
i_0\leq N-3, \nn\\ &&  \nn\\ &&  \sum_{k=1}^{[(N-1)/2]} (1-2k/N)r_k
~> j_2 + \sum_{[(N+1)/2]}^{N-4} ( 2k/N-1)r_k\ .\nn \eea
 Here are eliminated four chiral
generators  ~$X^+_{1,4+k}\,$, ~$X^+_{2,4+k}\,$,  $k\leq 2 \,$.

\eqna{zrsymnac} \buu{\bf ac}  && d ~=~  d^{ac} ~=~ 2 +j_1+j_2 + m_1\
, ~~~N\geq 4,\nn\\ && j_1 + m/N ~=~  j_2 + m^*/N  \ ,  \nn\\
&& j_1j_2>0, ~~ i_0+i'_0 =2 , \\
&& j_1>0, ~j_2=0, ~~ i_0=0, ~i'_0 =2, \\
&& j_1=0, ~j_2>0, ~~ i_0=2, ~i'_0 =0
  \ .\eena Here are eliminated four generators:  chiral generators ~$X^+_{1,4+k}\,$, $k\leq
1+i'_0(1-\d_{j_1,0})\,$, and anti-chiral generators
~$X^+_{3,5+N-k}\,$, $k\leq 1+i_0(1-\d_{j_2,0})\,$.

\eqnn{zrsymnad} \buu{\bf ad}  && d ~=~  d^{ad} ~=~ 1+j_2 + m_1 ~=~ 2m/N \ ,~~~N\geq 3 \ ,\\
&& j_1=0 , ~~  j_2 >0 , ~~i_0=1, ~~i'_0=0\ .  \nn \eea Here are
eliminated two chiral generators  ~$X^+_{1,5}\,$, $X^+_{2,5}\,$,
and two anti-chiral generators ~$X^+_{3,5+N-k}\,$, $k=1, 2
$.

\eqnn{zrsymnbc} \buu{\bf bc}  && d ~=~  d^{bc} ~=~ 1+j_1 + m_1 ~=~
2m^*/N \ ,~~~N\geq 3 \ ,\\ && j_2=0 , ~~~
   ~~j_1>0 , ~~i_0=0, ~~i'_0=1\ .  \nn \eea
Here   are eliminated  two chiral generators  ~$X^+_{1,4+k}\,$,
$k=1,2$, and two anti-chiral generators ~$X^+_{3,8}\,$,
~$X^+_{4,8}\,$.

\eqnn{zrsymnbd} \buu{\bf bd}  && d ~=~  d^{bd} ~=~ m_1 \ , ~~N\geq 2
\ ,\\  && j_1=j_2=0\  , ~~~~~i_0=i'_0=0\ .  \nn \eea   Here are
eliminated  two chiral generators  ~$X^+_{1,5}\,$, ~$X^+_{2,5}\,$,
and two anti-chiral generators ~$X^+_{3,8}\,$, ~$X^+_{4,8}\,$.

 Note that according to the results of   \cite{dob-bps}  the following cases would not be protected: ~{\bf ad}~ for
~$r_{N-1}>2$, ~{\bf bc}~ for ~$r_{1}>2$, ~{\bf bd}~ for
~$r_1,r_{N-1}>2$ when $N>2$, and for ~$r_1>4$ when $N=2$.

 \md

\section*{Acknowledgments}

The author would like to thank the Organizers for the kind
invitation to speak at the International Workshop 'Supersymmetries
and Quantum Symmetries', Dubna, 18-23.7.2011.\\ The author was
supported in part by Bulgarian NSF grant {\it DO 02-257}.

\end {document}